\documentclass[aps,prc,preprint,letter,floatfix,nofootinbib]{revtex4}

\usepackage{graphicx,epsfig,amsmath,amssymb}
\usepackage[bookmarksnumbered,bookmarksopen]{hyperref}
\usepackage[capitalize]{cleveref}
\usepackage[utf8]{inputenc}  
\usepackage[T1]{fontenc}  
\usepackage{listings}
\usepackage[normalem]{ulem}

\AtBeginDocument{\usepackage{booktabs}}
\usepackage{longtable}

\usepackage[usenames]{color}

\begin{document}

\setlength\LTcapwidth{\linewidth}

\title{Constraining resonance properties through kaon production in pion-nucleus collisions at low energies}

\author{V.~Steinberg$^1$, J.~Steinheimer$^1$, H.~Elfner$^{2,3,1}$, M.~Bleicher$^{1,2,3,4}$}

\affiliation{$^1$Frankfurt Institute for Advanced Studies, Ruth-Moufang-Strasse 1, 60438 Frankfurt am Main, Germany}
\affiliation{$^2$GSI Helmholtzzentrum für Schwerionenforschung, Planckstr. 1, 64291 Darmstadt, Germany}
\affiliation{$^3$Institute for Theoretical Physics, Goethe University, Max-von-Laue-Strasse 1, 60438 Frankfurt am Main, Germany}
\affiliation{$^4$John von Neumann-Institut für Computing, Forschungszentrum Jülich, 52425 Jülich, Germany}
\date{\today}

\begin{abstract}
Hadronic interactions are crucial for the dynamical description of heavy-ion reactions at low collision energies and in the late dilute stages at high collision energies.
In particular, the properties and decay channels of resonances are an essential ingredient of hadronic transport approaches.
The HADES collaboration measured particle production in collisions of pions with carbon and tungsten nuclei at $E_\text{kin} = 1.7\,\text{GeV}$~\cite{Adamczewski-Musch:2018eik}.
Such reactions are of high interest, because they allow to probe the properties of baryonic resonances produced in a much cleaner environment than the usual nucleus-nucleus collisions.
We study these reactions with two transport approaches: SMASH (Simulating Many Accelerated Strongly-interacting Hadrons) and UrQMD (Ultra relativistic Quantum Molecular Dynamics) which follow the same underlying concept but with different implementations.
The differential spectra in rapidity and transverse momentum are used to show how model parameters, as the decay channels of high mass resonances and angular distributions of kaon elastic scattering, can be constrained. It is found that the data favor the production of more particles with lower momentum over the production of few particles with higher momentum in these decays. 
In addition, the shape of the rapidity distribution of the kaons strongly depends on the angular distribution of the elastic kaon-nucleon cross section.
\end{abstract}

\maketitle


\section{Introduction}

The production of strange hadrons in heavy-ion collisions provides an important probe to extract the properties of dense nuclear matter~\cite{Koch:1986ud,Aichelin:1986ss}.
Because strange hadrons are not present in ordinary matter, they have to be newly produced.
To study the production of strangeness in dense and hot systems is a goal of current and future experiments such as at the Gesellschaft für Schwerionenforschung~(GSI)~\cite{Wilczek:2010ae}/Facility for Antiproton and Ion Research~(FAIR)~\cite{Ablyazimov:2017guv}, the Nuclotron-based Ion Collider~(NICA)~\cite{Kekelidze:2012zz}, the Japan Proton Accelerator Research Complex~(J-PARC)~\cite{Sako:2014fha}, and the Relativistic Heavy-Ion Collider~(RHIC)~\cite{Aggarwal:2010cw}.

Since ab-initio calculations of quantum chromodynamics are currently not possible for the high baryon densities of these systems~\cite{Allton:2002zi,Kaczmarek:2011zz,Borsanyi:2012cr,Borsanyi:2013hza,Gavai:2008zr}, phenomenological models are required to extract the properties of dense matter by comparing model results to experimental data.
Usually, hadronic transport approaches such as IQMD~\cite{Hartnack:1997ez}, UrQMD~\cite{Bass:1998ca}, HSD~\cite{Cassing:1999es}, JAM~\cite{Nara:1999dz}, GiBUU~\cite{Buss:2011mx} and SMASH~\cite{Weil:2016zrk} are employed.
At low energies, the models GiBUU~\cite{Agakishiev:2014moo}, IQMD and HSD~\cite{Hartnack:2011cn} describe strangeness production by parametrizing the direct cross sections and employing kaon-nucleon and antikaon-nucleon potentials, while UrQMD~\cite{Steinheimer:2015sha} and SMASH~\cite{smash_strangeness} include high-mass nucleon resonances for that purpose.

With regard to the properties of heavy resonances, the recent measurements of the momentum spectra of kaons and $\Lambda$~baryons in $\pi^-$-$C$ and $\pi^-$-$W$ collisions~\cite{Adamczewski-Musch:2018eik} at $E_\text{kin} = 1.7\,\text{GeV}$ by the HADES collaboration are of great interest.
This system is particularly suited to constrain the properties of heavy resonances, because the $\pi N \to N^*, \Delta^*$ channel is the dominant inelastic channel.
In low-energy nucleus-nucleus collisions, the $NN \to NN^*$ channel dominates baryonic resonance production.
Pion-nucleus reactions offer the intriguing possibility to excite prominently high mass resoances, because the center-of-mass energy can be finely adjusted (in contrast to nucleus-nucleus reactions).

In this work, it is investigated how resonance properties (mainly branching fractions) determine the production of $K^+$ in the pion beam within a hadronic transport approach.
It is particularly beneficial to compare the results from two independent implementations with similar physics assumptions, to demonstrate which results are independent of the particular model in use.
To this effect, the models SMASH and UrQMD are studied in the following.

The paper is divided in three parts:
First, the models are described, in particular the production process of strangeness. The total cross section of the pion-nucleus interaction is given as well.\\
Second, we present the transverse momentum spectra of $K^+$ for the pion beam and show how they constrain some properties of heavy resonances such as the $N^*, \Delta^* \to KY, KY\pi$~decays (where $Y$ refers to hyperons), which are not well constraint by other experimental measures.\\
Third, the rapidity spectra of $K^+$ are investigated and shown to be sensitive to the, mostly unmeasured, angular distribution of the elastic kaon-baryon cross section.

\section{Model description}
\label{sec:models}
\subsection{SMASH}
\label{sec:smash}

SMASH~\cite{Weil:2016zrk} is a recently introduced hadronic transport approach, incorporating the newest available experimental data~\cite{Tanabashi:2018oca,Munzer:2017hbl,Agakishiev:2012vj} to provide a baseline at low energies that can be extended with the physics required by intermediate energies.

The strangeness production in SMASH was tuned to elementary cross sections~\cite{smash_strangeness}:
\begin{enumerate}
\item The individual $N^*,\, \Delta^* \to YK, YK^*$ branching ratios were measured for some resonances~\cite{Tanabashi:2018oca}, providing upper and lower bounds. However, the uncertainty is often large, especially for heavy resonances, where sometimes the relevant channels were never measured.
In particular, there are some measurements of $N^* \to \Lambda K$~branching ratios, but not for $N^* \to \Sigma K$.
\item Some of the relevant exclusive cross sections were measured~\cite{landolt_boernstein}, constraining the total contribution of the resonances:
\begin{itemize}
    \item $p\pi \to YK$
    \item $pp \to KX, \Lambda X, YNK$
\end{itemize}
See \cite{smash_strangeness} for a more detailed discussion.
Unfortunately, there is no data on cross sections with $YK\pi$ in the final state to constrain the possible $N^*, \Delta^* \to Y K^*$ branching ratios.
\end{enumerate}

SMASH~1.3 was compared to the HADES measurements of kaon production in proton-proton collisions at $E_\text{kin} = 3.5\,\text{GeV}$, nickel-nickel collisions and gold-gold collisions at $E_\text{kin} = 1.5A\,\text{GeV}$ and Ar-KCl collisions at $1.76A\,\text{GeV}$~\cite{smash_strangeness}.
The model was able to roughly describe the nickel-nickel and Ar-KCl data.
There was also a good agreement with the gold-gold data for small numbers of participants, but for large participant numbers or high rapidities, the kaon and antikaon production was overestimated.

Also, the dilepton production observed by HADES at~$E_\text{kin} = 1-3.5A\,\text{GeV}$ in nucleon-nucleon, proton-niobium, carbon-carbon and Ar-KCl was studied with SMASH~\cite{Staudenmaier:2017vtq}, which was found to be in good agreement with the data.
In particular, the $\phi$ production in SMASH was constrained using the HADES data on the invariant mass spectra of dielectrons produced in proton-niobium collisions.

Another important ingredient for modeling kaons in the medium is the angular distribution of the elastic kaon-nucleon scattering. The treatment of this cross section in fact depends on the version employed.

In SMASH~1.301, all cross sections are isotropic, except for $NN \to NN, N\Delta$, which use the Cugnon parametrization~\cite{Cugnon:1996kh}. For proton-proton, this is given by
\begin{equation}
\frac{d\sigma}{dt} \propto \exp(Bt)
\; ,
\end{equation}
where $t$ is the usual Mandelstam variable and the coefficient~$B$ depends on the incident lab momentum~$p_\text{lab}$ in units of $\text{GeV}$:
\begin{equation}
B = \begin{cases}
5.5 p_\text{lab}^8/(7.7 + p_\text{lab}^8) & p_\text{lab} < 2 \\
5.334 + 0.67 (p_\text{lab} - 2) & \text{otherwise}
\end{cases}
\label{eqn:cugnon_coeff}
\end{equation}
In SMASH~1.6, this parametrization is employed for all elastic scattering, including the kaon-nucleon scattering.

Additional changes from SMASH~1.301 to 1.6 (which corresponds to the version discussed in~\cite{smash_strangeness}) are the following:
\begin{itemize}
\item The $N^*, \Delta^* \to YK$ decays we reduced in favor of $N^*, \Delta^* \to YK^*$ decays. See \cref{sec:urqmd,sec:pt} for a motivation of this change.
\item SMASH~1.6 uses string fragmentation. This affects the $\pi N$ interaction at $1.9\,\text{GeV} \le \sqrt s \le 2.2\,\text{GeV}$, where SMASH interpolates between resonances and strings. At higher energies, strings are used exclusively. However, production of strangeness via resonances is exempt from this, because there is no strangeness production from strings at these energies in SMASH.  
\item The angular momenta of decays were updated to respect parity and angular momentum conservation.
\item The particle properties were updated to the PDG data from 2018~\cite{Tanabashi:2018oca}. This includes many changes: The masses and widths of the particles were updated, one~$\Delta^*$ and five~$N^*$ resonances were added, and the branching ratios and matrix elements were retuned to the elementary cross sections.
\end{itemize}
For more details on the particle properties, see \cite{smash_strangeness} for SMASH~1.301 and \cite{smash16} for SMASH~1.6\footnote{The particle properties are defined in the files {\tt input/particles.txt} and {\tt input/decaymodes.txt}.}.

\subsection{UrQMD}
\label{sec:urqmd}

The UrQMD model is a microscopic transport description for nuclear collisions at various beam energies, ranging from GSI-SIS to the LHC. 
Particles are either propagated in the nuclear mean field or on straightline trajectories, if the cascade mode is chosen.
Particles scatter based on their cross sections.
Here, elastic and inelastic reactions are possible.
At moderate collision energies, string formation and decays are treated as well.
The UrQMD model is extensively described in Refs.~\cite{Bass:1998ca,Bleicher:1999xi,Steinheimer:2015sha}.
In the energy range considered in this paper, i.e. $E_\text{c.m.} \approx 2\,\text{GeV}$, the interactions between the incoming meson and the nucleons is dominated by resonance excitations and strings only play a minor role.
Thus, the production of new particles at HADES energies, like the kaons, dominantly goes through the excitation and decay of a heavy baryonic resonance. Here the probability of producing such a state as well as the probability that such a state decays into a kaon and associated hyperon, is determined by the so-called branching ratios.
These have been taken from the PDG~\cite{Tanabashi:2018oca} and, in part, extracted from elementary reactions.
Unfortunately, for the high mass resonances relevant in the present paper, many of these branching ratios are not well constrained by the presently available data.
UrQMD has since long been used to explore the HADES energy regime.
See e.g. \cite{Endres:2015fna,Schmidt:2008hm} for dilepton studies and \cite{Hillmann:2018nmd,Hillmann:2019wlt} for flow studies, while in-medium effects were explored e.g. in \cite{Reichert:2019lny}.

\cref{tab:br_ur} gives a summary of the branching ratios implemented in the current version of the UrQMD model, version 3.4. As one can see, the high mass resonances show a large direct decay probability in the hyperon+kaon channel, which will leave the produced kaons with a rather large momentum. To study the effect of these relatively unknown branching ratios, we will compare our results with a modified version of UrQMD (dubbed UrQMD~\textit{mod.} for the present paper) where the branching ratio to the hyperon+kaon final state are reduced. In order to keep the kaon production cross section fixed, the resonance decays via the chain hyperon+$K^*$ and $K^* \to K\pi$, thus only the final momentum of the kaon gets modified.

\begin{table}[t]
\addtolength{\tabcolsep}{0.3em}
\begin{tabular}{ccccc}
\toprule 
\scriptsize	  Ratio $[\%]$ &\multicolumn{2}{c}{ \scriptsize $\Gamma_{\Lambda K}/\Gamma_{tot} $} & \multicolumn{2}{c}{ \scriptsize$\Gamma_{\Sigma K}/\Gamma_{tot} $} \\ \midrule	
\midrule	
\scriptsize 	Resonance           &	  v3.4  &  UrQMD \textit{mod.}  &  v3.4  & UrQMD \textit{mod.}  \\ \midrule
\scriptsize   N*(1650) &	 7  &  8    &  2  &  0  \\
\scriptsize   N*(1710) &	10  &  8    &  3  &  0  \\
\scriptsize   N*(1720) &	10  &  7    &  2  &  0.5  \\
\scriptsize   N*(1900) &	2   &  0.5  &  0  &  0.5  \\
\scriptsize   N*(1990) &	3   &  3    &  0  &  0  \\
\scriptsize   N*(2080) &   12   &  0    &  0  &  0  \\
\scriptsize   N*(2190) &	12  &  0  &  0  &  0  \\
\scriptsize   N*(2220) &	12  &  0  &  0  &  0  \\
\scriptsize   N*(2250) &	12  &  0  &  0  &  0  \\
\scriptsize   $\Delta(1920)$ &	 0  &  0  &  3 &  3  \\
\scriptsize   $\Delta(1930)$ &	 0  &  0  & 15 &  0  \\
\scriptsize   $\Delta(1950)$ &	 0  &  0  & 12  &  0  \\
\bottomrule
\end{tabular}
\addtolength{\tabcolsep}{-0.3em}
\caption{Branching ratios of heavy baryonic resonances to $\Lambda+K$ and $\Sigma + K$ in UrQMD~3.4 and compared to a modified version (UrQMD \textit{mod.}), described in the text.}
\label{tab:br_ur}
\end{table}

Another input in the model simulations that influences the transport cross section of kaons in the medium, is the angular distribution of the meson+baryon elastic scattering. For the scattering of kaons with nucleons this is not well known and in particular in the dense medium of a nucleus this angular distribution may change. In the following study we will implement two different angular distributions for the kaon+nucleon elastic scattering. The probability $P(\theta)$ that the particles are scattered into a given angle~$\theta$ w.r.t. to the incoming momenta, in the center of mass frame, is given as:

\begin{equation}
P(\theta)\propto \exp{(a \theta)}   \ . 
\end{equation}
In the standard version of UrQMD the parameter $a$ is set to $a=8$, which gives a rather forward peaked scattering angle distribution. If $a$ is set to $a=0$, the distribution will be isotropic, a scenario which will be studied later.

\section{Total cross section}

\begin{figure*}
\centering
\includegraphics[width=0.49\linewidth]{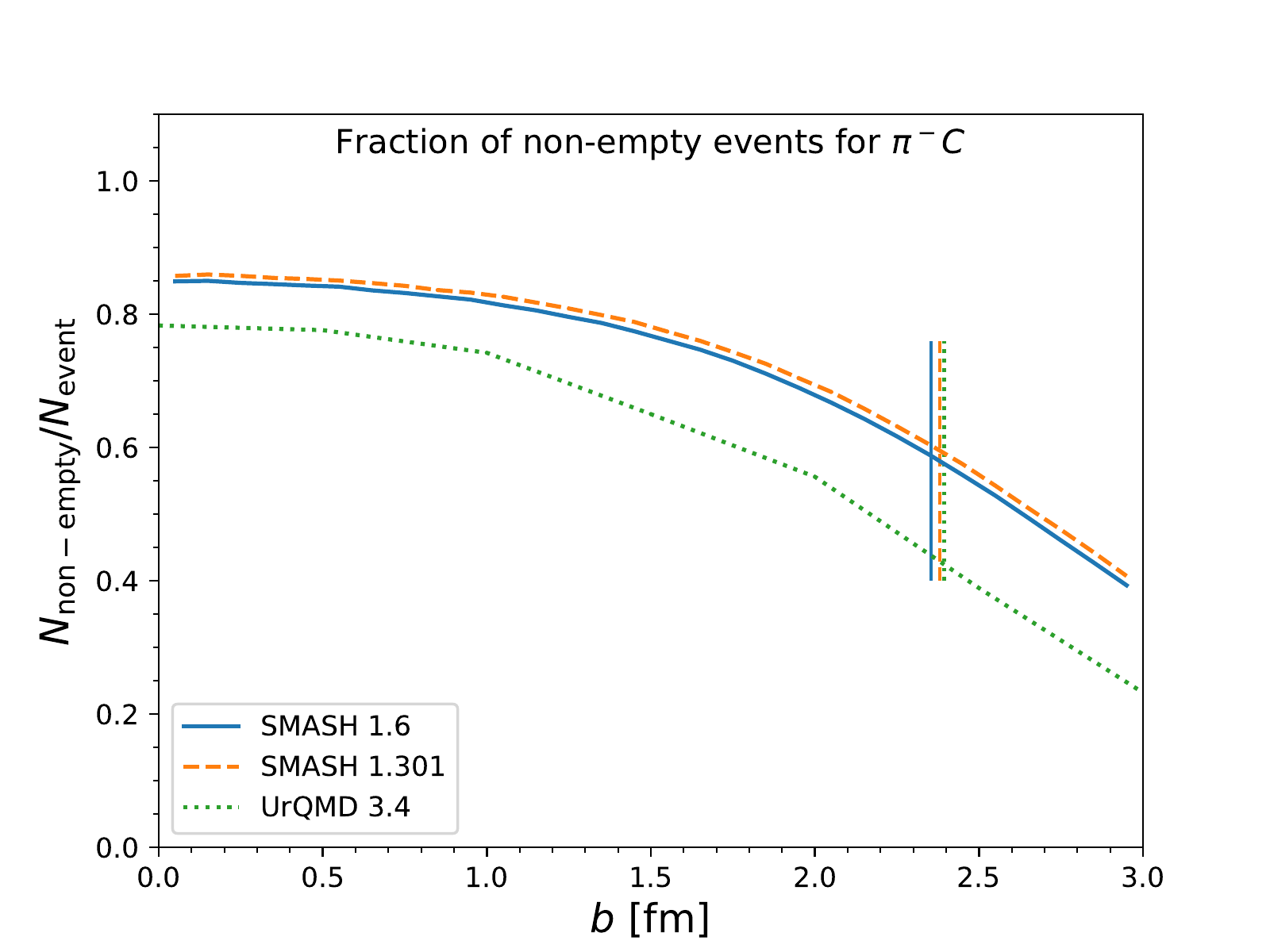}
\includegraphics[width=0.49\linewidth]{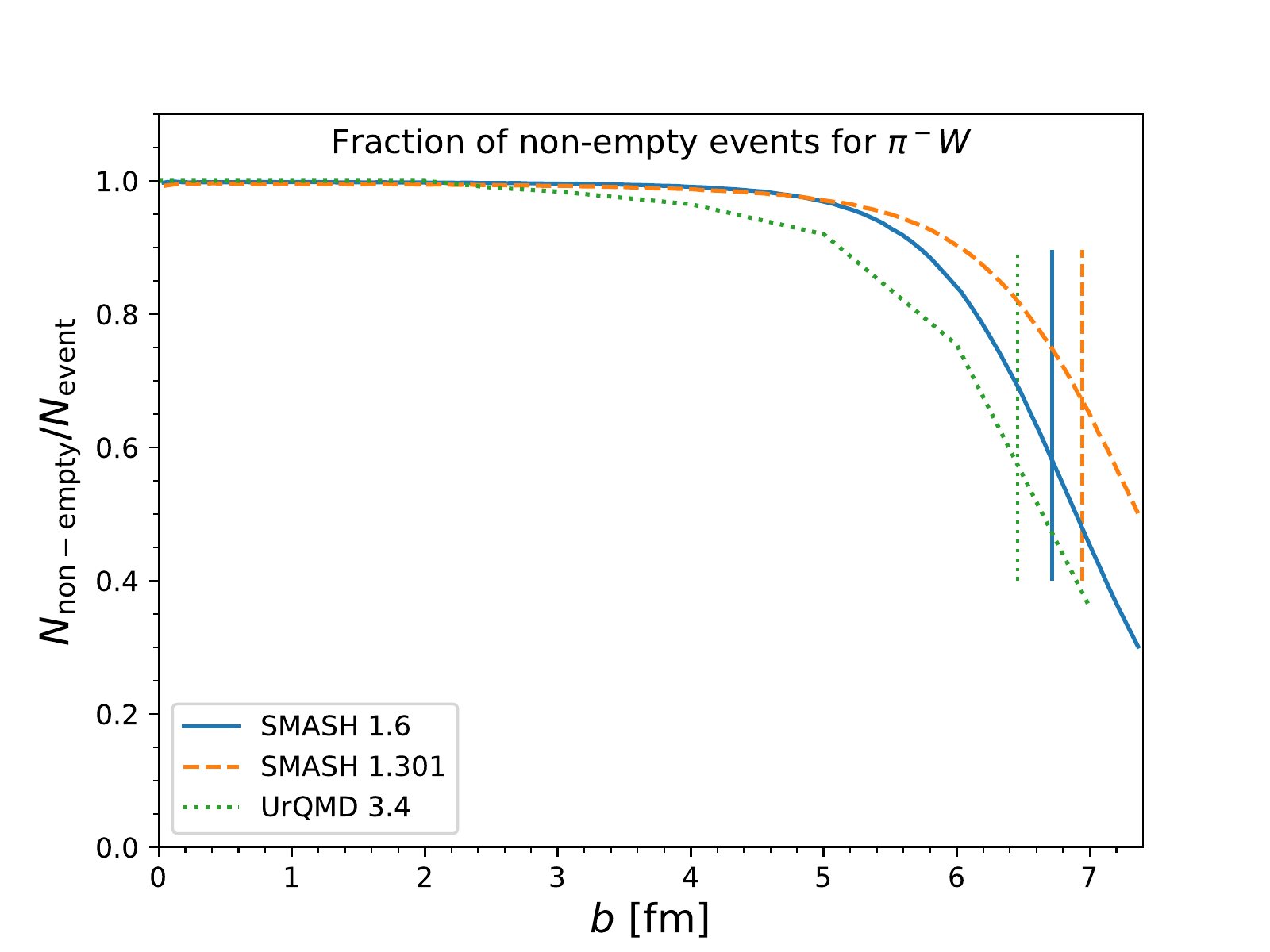}
\caption{
Fraction of non-empty events~$N_\text{non-empty}/N_\text{total}$ for SMASH~1.6 (continuous line), SMASH~1.301 (dashed line) and UrQMD~3.4 (dotted line) in $\pi^-$-$C$ (left) and $\pi^-$-$W$ (right).
The vertical lines give the value of~$b_\text{eff}$, which corresponds to the pion-nucleus cross section as defined by~\cref{eqn:total_xs} and given in \cref{tab:total_xs}.
}
\label{fig:impact}
\end{figure*}

For the simulations of the collision of a $\pi^-$~beam at $E_\text{kin} = 1.7\,\text{GeV}$ with a carbon and a tungsten target, the impact parameter~$b$ was sampled from a minimum-bias distribution.
In order to compare to the experimental data of HADES, it is necessary to determine the total pion-nucleus cross section to normalize the momentum spectra.
Assuming a geometrical overlap geometry, i.e. assuming the nucleus as a homogeneous sphere with radius~$b_\text{eff}$, the total cross section is given by
\begin{equation}
    \sigma_\text{tot}
    = \pi b_\text{eff}^2
    = \pi b_\text{max}^2 \frac{N_\text{non-empty}}{N_\text{total}}
    \;,
\label{eqn:total_xs}
\end{equation}
where $b_\text{max}$~is the maximal sampled impact parameter, $N_\text{non-empty}$~is the number of events where the pion collides with a nucleon and $N_\text{total}$~is the total number of events.
The fraction of non-empty events as a function of impact parameter~$b$ is shown in \cref{fig:impact}.
It can be seen that it does not fall off sharply at~$b_\text{eff}$ as would be expected if the nucleus was a sphere, but rather drops smoothly.
This is because in the hadronic transport calcuations employed here, the nucleus consists of nucleons randomly sampled from an assumed nucleon distribution: For high impact parameters there is only a finite probability for the projectile to hit a nucleon from the target, resulting in a systematic uncertainty when calculating the total pion-nucleus cross section.
For small nuclei such as carbon this uncertainty is larger, because carbon has only few nucleons and becomes "transparent" to the pion even for small impact parameters.

The resulting total cross sections of the pion and the carbon and tungsten target for SMASH and UrQMD are listed in \cref{tab:total_xs}, which differ by less than 10\% between the two models.

\begin{table}
    \centering
    \begin{tabular}{cccc}
         \toprule
         System & \multicolumn{3}{c}{$\sigma_\text{tot}$} \\
         & SMASH 1.6 & SMASH 1.301 & UrQMD 3.4 \\
         \midrule
         $\pi^- C$ & 180 mb & 184 mb & 186 mb  \\
         $\pi^- W$ & 1432 mb & 1530 mb & 1310 mb \\
         \bottomrule
    \end{tabular}
    \caption{Total cross section~$\sigma_\text{tot}$ for the pion-nucleus interaction of the given systems with SMASH and UrQMD.}
    \label{tab:total_xs}
\end{table}

\subsection{Kaon production cross section}
\label{sec:kaon_prod}

First, we present the integrated mean multiplicity of $K^+$ for all the different model implementations: SMASH~1.301, SMASH~1.6, UrQMD~3.4 and UrQMD~\textit{mod.} As can be seen in \cref{tab:kaon_mult}, the four models yield up to a factor of~6 different numbers of produced kaons.
This is a direct result of the many mainly unknown resonance properties at the center-of-mass energy of the interactions studied.
The large difference (factor~6) in the kaon yield between SMASH~1.301 and SMASH~1.6 is due to modified branching ratios as discussed at the end of \cref{sec:pt}.
Consequently, this means that the HADES pion beam data can be used to gauge the model parameters for resonances and the onset of string formation in this particular energy region.

\begin{table}[b]
    \centering
    \begin{tabular}{cccc}
         \toprule
 \multicolumn{4}{c}{$K^+$ multiplicity} \\
         \midrule
         \midrule
 \multicolumn{4}{c}{$\pi^- C$ } \\
         \midrule         
         SMASH 1.6 & SMASH 1.301 & UrQMD \textit{mod.} & UrQMD 3.4  \\
         \midrule
         0.018 & 0.0028 & 0.0065 & 0.0087 \\
         \midrule
         \midrule
 \multicolumn{4}{c}{$\pi^- W$ } \\
         \midrule   
         SMASH 1.6 & SMASH 1.301 & UrQMD \textit{mod.} & UrQMD 3.4  \\
         \midrule
         0.022 & 0.0051 & 0.0066 &  0.0089 \\
         \bottomrule         
    \end{tabular}
    \caption{Average multiplicity of positively charged kaons in the different model implementations and reactions. The results are for the minimum bias collisions as described in the previous section.}
    \label{tab:kaon_mult}
\end{table}

In the following we will investigate in more detail the shapes of the transverse momentum and rapidity distribution of positively charged kaons in the different model implementations.
Here, we will focus on the shapes, because the total kaon yield is not well constrained in these reactions as was discussed above.

\section{Transverse momentum spectra}
\label{sec:pt}
\begin{figure*}[t]
\centering
\includegraphics[width=0.49\linewidth]{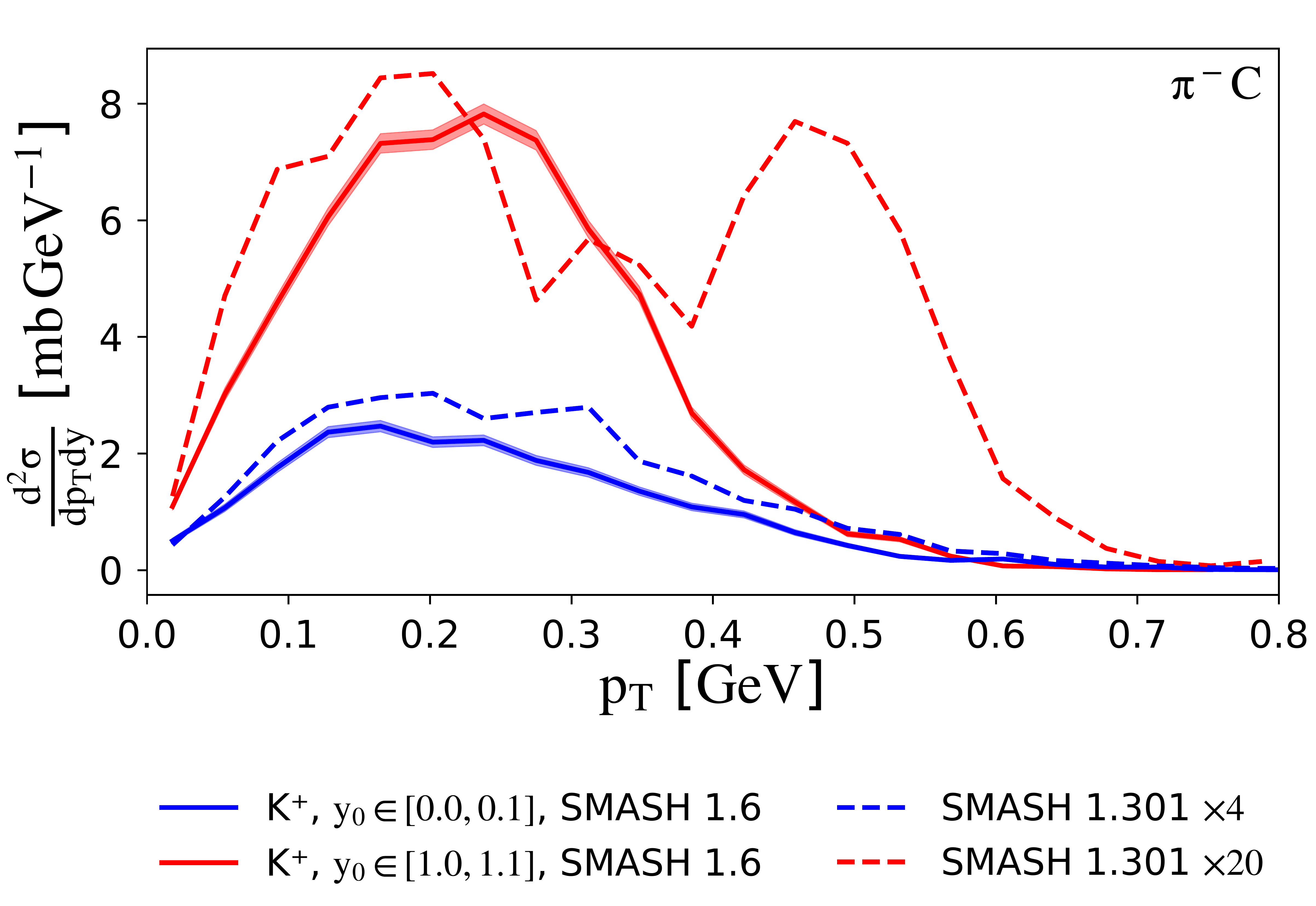}
\includegraphics[width=0.49\linewidth]{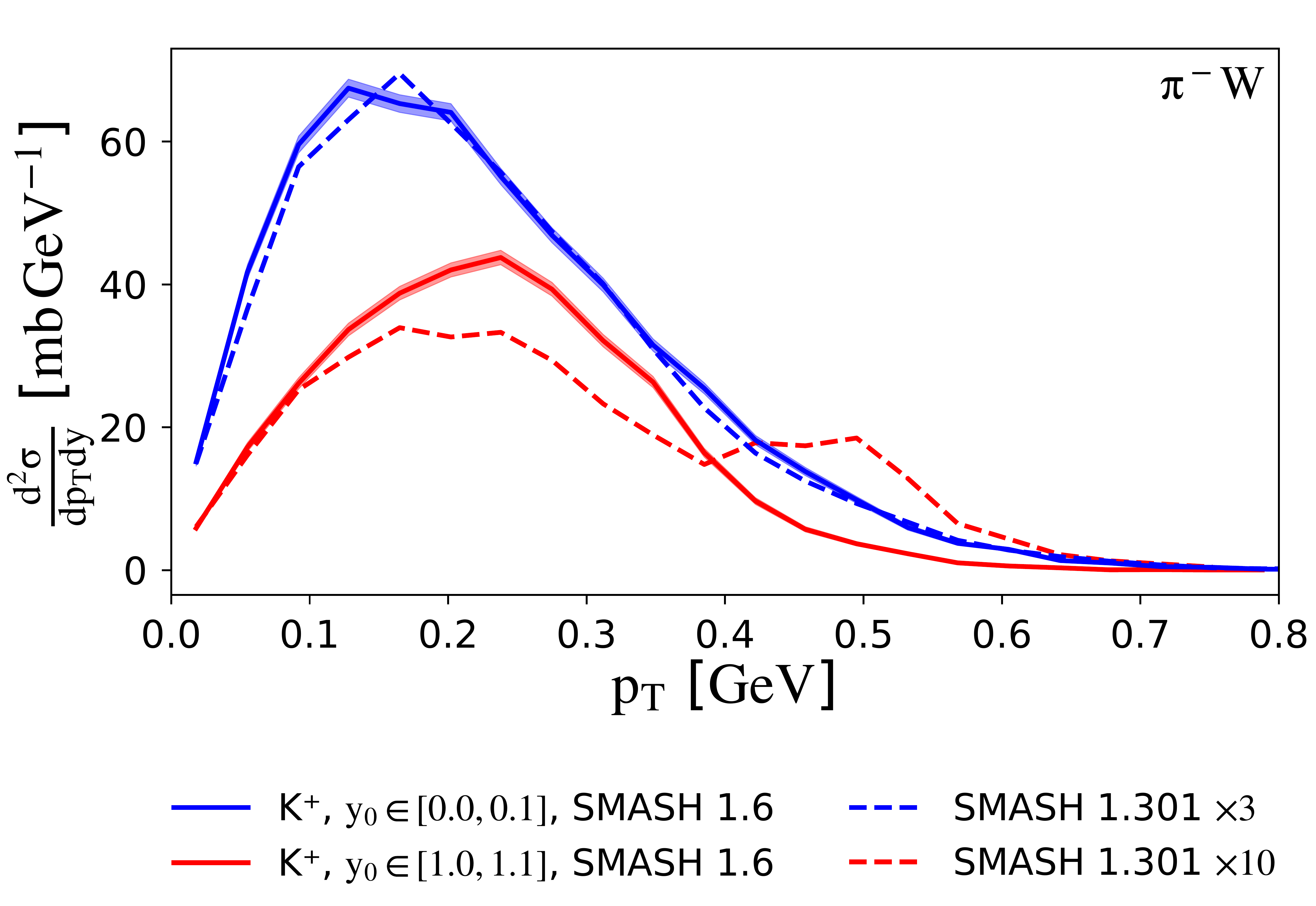}
\includegraphics[width=0.49\linewidth]{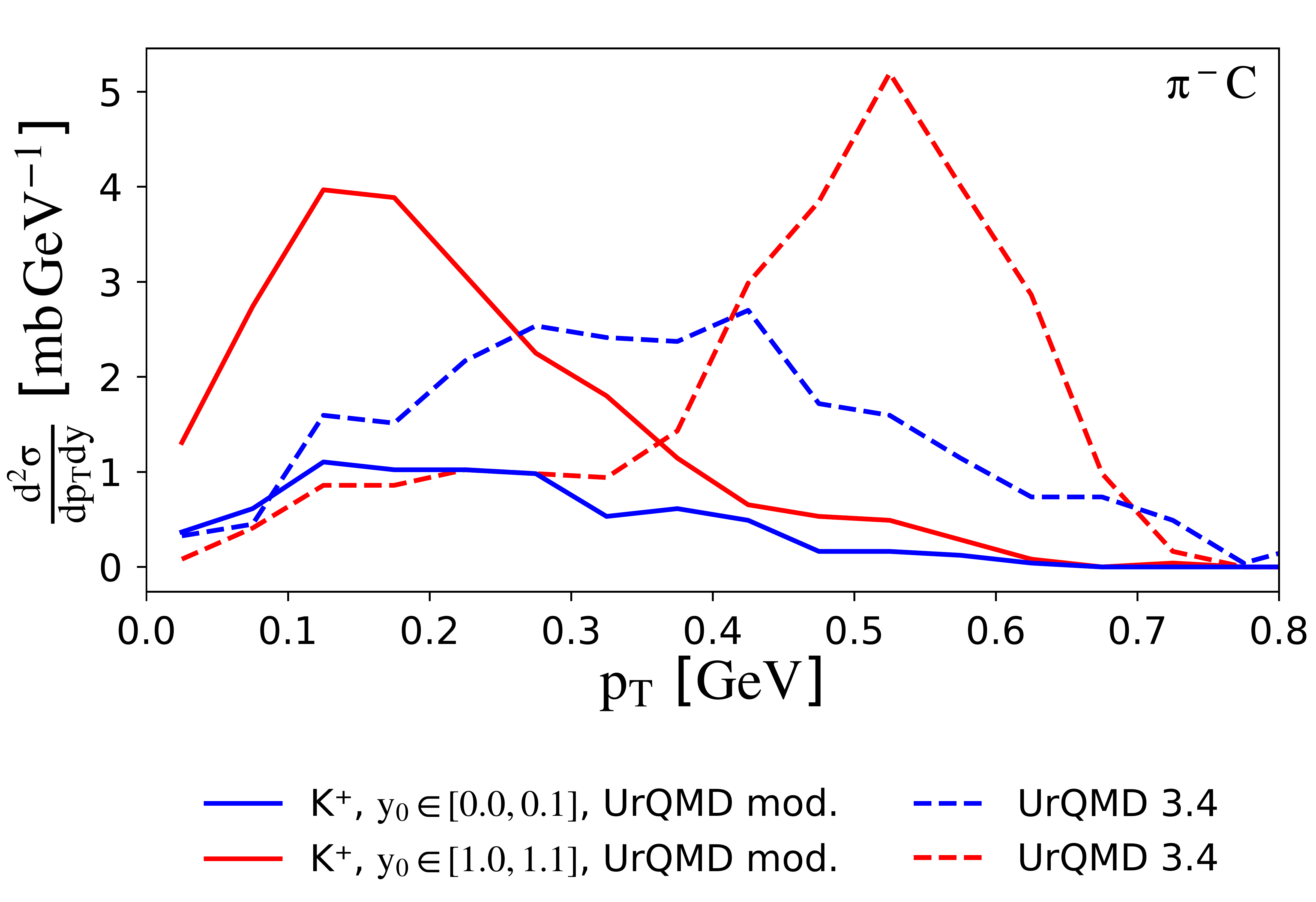}
\includegraphics[width=0.49\linewidth]{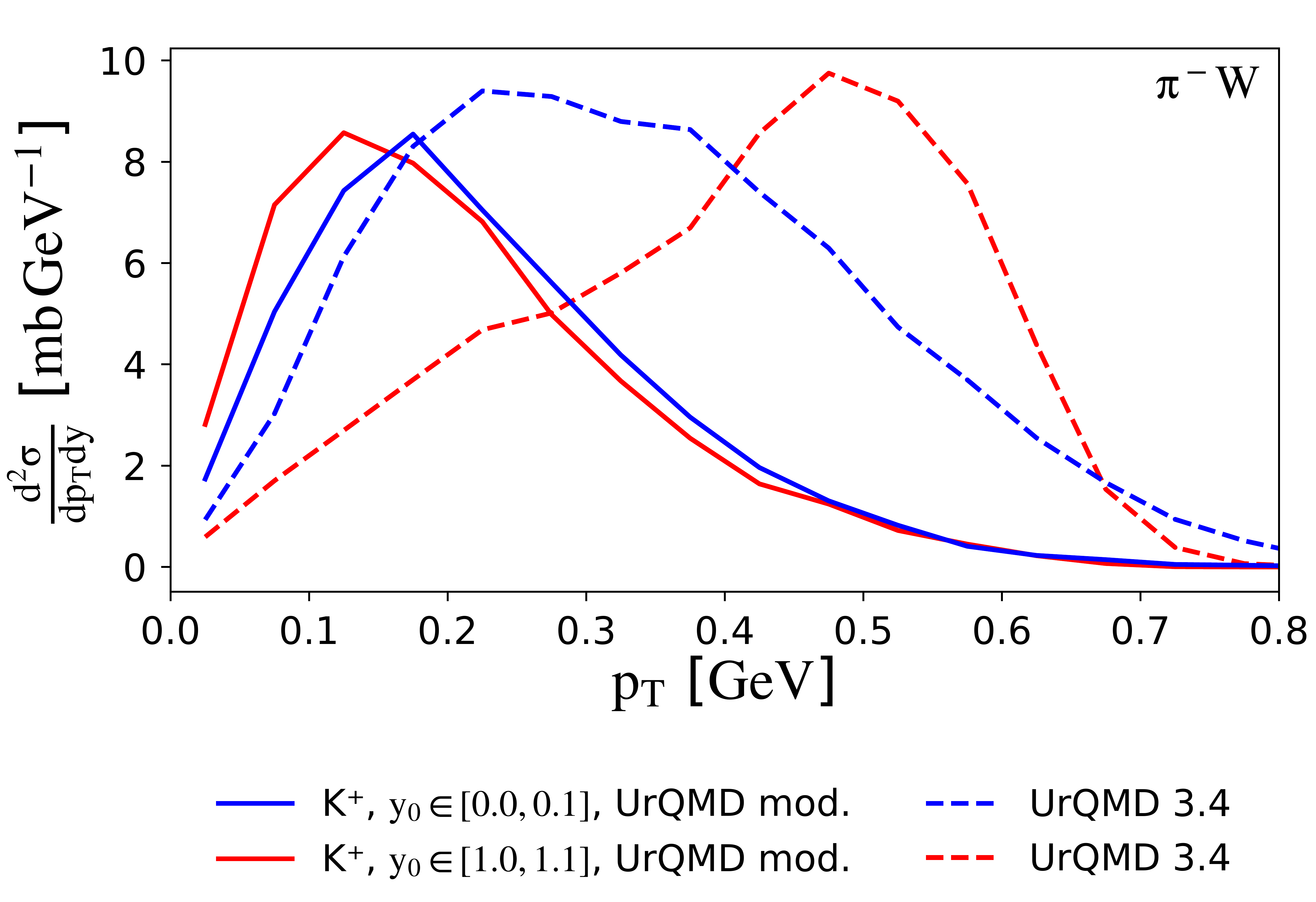}
\caption{$K^+$ transverse momentum spectra for SMASH~(top) and UrQMD~(bottom), in $\pi^+$-$C$ (left) and $\pi^+$-$W$ (right), normalized by the total pion-nucleus cross section given in \cref{tab:total_xs}. The dashed lines show SMASH~1.3 and UrQMD~3.4 results with a peak at high~$p_T$ from heavy resonance decays, while the continuous lines show SMASH~1.6 and UrQMD~3.4 with modified branching ratios. To allow for an easier comparison of the shape, some of the SMASH results have been multiplied with factors given in the legend.}
\label{fig:pt_kplus}
\end{figure*}

\cref{fig:pt_kplus} shows the transverse momentum spectra in hte $\pi+$ and $\pi+W$ systems for selected rapidities, normalized by the total cross section~$\sigma_\text{tot}$.
For SMASH~1.301 and UrQMD~3.4, it can be seen in \cref{fig:pt_kplus} that at low rapidities ($y \in [0, 0.1]$), the spectrum has a single peak at low $p_T \approx 0.2-0.3\,\text{GeV}$.
However, for high rapidities~$y \in [1.0, 1.1]$, there is a second peak at high $p_T \approx 0.5\,\text{GeV}$ appearing in both approaches.
The second peak can be understood by looking at how the $K^+$ are produced in pion-nucleon collisions:
As discussed above, their production is dominated by $N^*,\, \Delta^*$~decays into kaons, hyperons and possibly pions. In particular, the direct two-particle decays~$N^*,\, \Delta^* \to YK$ have the highest branching ratios, even for heavy resonances.
This results in high momentum kaons from the decays of resonances, resulting in the second peak at high rapidities.

This indicates that high mass resonances predominantly decay via chains and not directly to ground state hadrons.
As a result, the additional particle carries energy away and the kaon is produced with lower transverse momentum~$p_T$.
This argument is supported by the experimental results~\cite{Adamczewski-Musch:2018eik}, where the transverse momentum distributions are peaked at lower momenta ($p_T \approx 0.2\,\text{GeV}$), corresponding to slower kaons.
To achieve this in the models, the direct $N^*,\, \Delta^* \to YK$ branching ratios of heavy resonances can be decreased in favor of increasing or introducing $N^*,\, \Delta^* \to YK^*$ branching ratios of similar weight, which leads to the creation of an additional pion in the final state.

The branching ratios of heavy resonances have been adapted in both approaches as described in \cref{sec:models} above.
The modified versions are dubbed as SMASH~1.6 and UrQMD~\textit{mod.}. The secondary peak at high $p_T \approx 0.5$ is essentially removed from the $K^+$~spectra if the decay chain dominates for high mass resonances (see~\cref{fig:pt_kplus}, continuous lines).

As seen in~\cref{fig:pt_kplus}, SMASH~1.6 produces significantly more kaons than SMASH~1.301.
While both versions describe the cross sections listed above within the experimental uncertainty, the strangeness-production cross sections in SMASH~1.6 are substantially higher.
More significantly, SMASH~1.6 contains many new $N^*, \Delta^* \to YK^*$ channels which are absent in SMASH~1.301, increasing kaon production.
As discussed above, these channels are barely constrained by experimental data and can be modified once tighter experimental constraints become available.

\section{Rapidity spectra}
\label{sec:rapidity}

\begin{figure*}[t]
\centering
\includegraphics[width=0.49\linewidth]{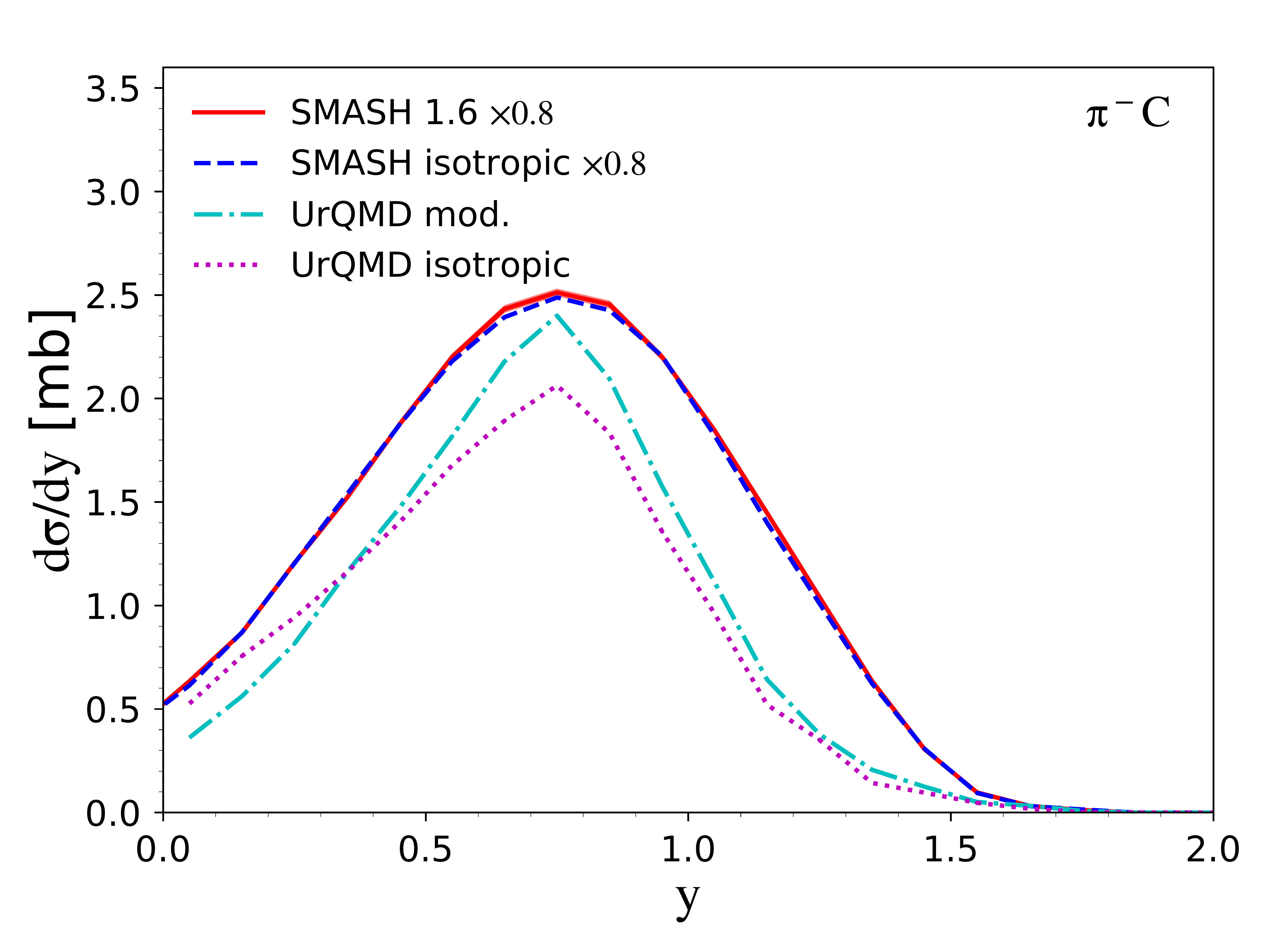}
\includegraphics[width=0.49\linewidth]{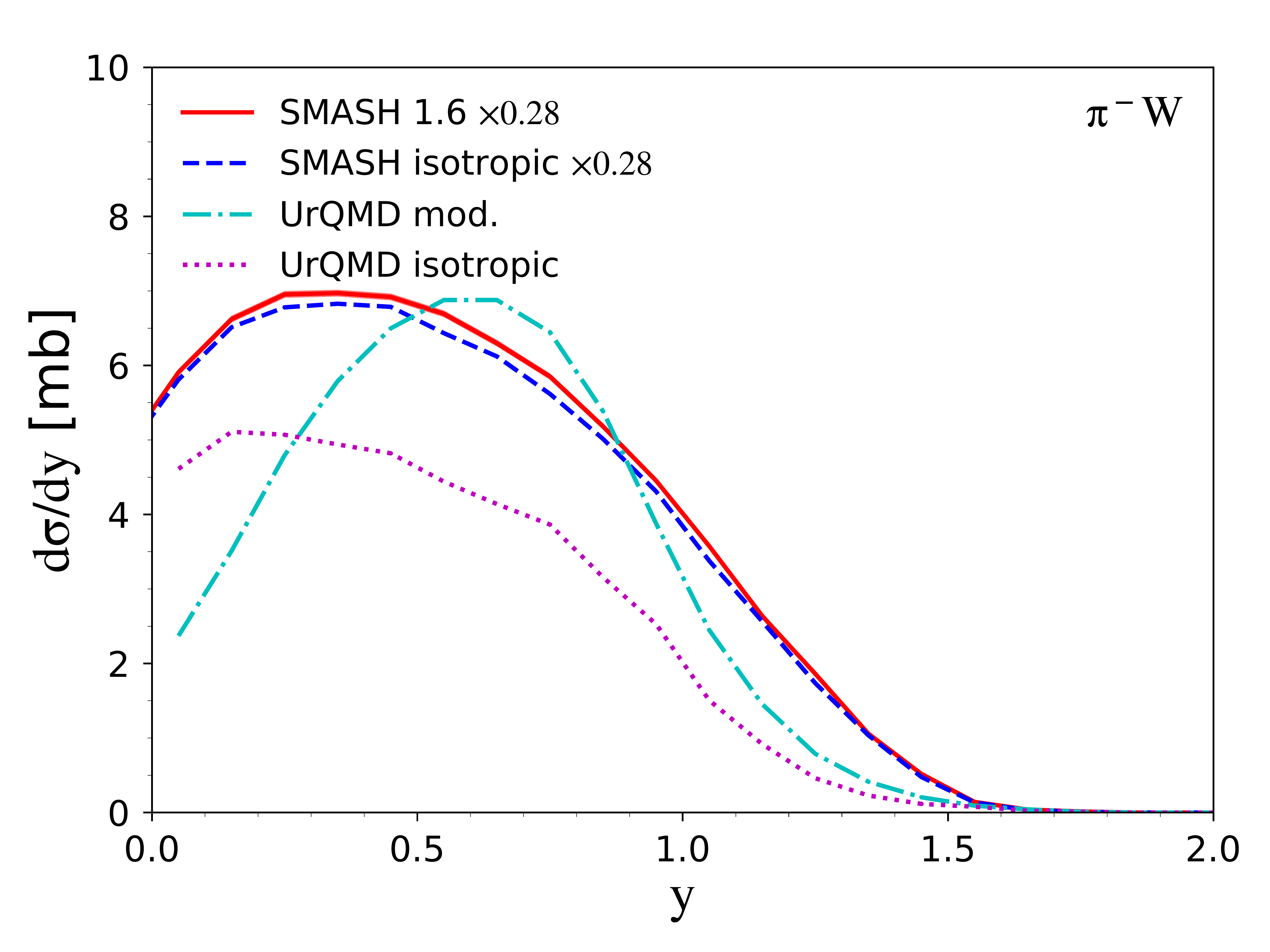}
\caption{
$K^+$ rapidity spectra for SMASH~1.6 and UrQMD~\textit{mod.} in $\pi^-$-$C$ (left) and $\pi^-$-$W$ (right), normalized by the total pion-nucleus cross section given in \cref{tab:total_xs}.
For both models, modified versions with isotropic elastic cross sections are also shown.
To allow for an easier comparison, the SMASH results were scaled down by the factor given in the legend.}
\label{fig:y_kplus}
\end{figure*}

\cref{fig:y_kplus} shows the rapidity spectra for the same systems as in \cref{sec:pt}, again normalized by the total cross sections~$\sigma_\text{tot}$ from \cref{tab:total_xs}.
As discussed in \cref{sec:kaon_prod}, the total $K^+$ yield for both models is very different.
Consequently, the following discussion focuses on the shape of the spectra, and the plots of the spectra are rescaled for visibility.

For the carbon target, the position of the peak is similar for both models ($y \approx 0.7$).
This means that the kaons are essentially produced at midrapidity of the $\pi + p$~system ($y_\text{mid} \approx 0.8$).
However, for the tungsten target, the peaks are centered at different rapidities for SMASH~1.6 ($y \approx 0.3$) and UrQMD~\textit{mod.} ($y \approx 0.6$).
Both differ from the HADES results~\cite{Adamczewski-Musch:2018eik}, where the peak is centered at $y \approx 0$.

This difference can be understood as follows:
For the tungsten target, an elastic rescattering of the produced $K^+$ with other nucleons is much more likely than for the smaller carbon target.
Thus, the position of the peak in the rapidity spectrum depends on the angular distribution of this scattering.
In SMASH~1.6, all elastic scatterings are anisotropic with an angular distribution given by the Cugnon parametrization of nucleon-nucleon cross sections (see \cref{sec:smash}), while UrQMD~3.4 and UrQMD~\textit{mod.} use a different anisotropic parametrization which can be considered as mainly forward-backward peaked (see \cref{sec:urqmd}).
Changing UrQMD~\textit{mod.} to use isotropic meson-baryon scattering moves the peak to lower rapidities, while for SMASH~1.6 there is no significant effect (see \cref{fig:y_kplus}), because the Cugnon parametrization is already almost isotropic for the low momenta of the kaons (see \cref{eqn:cugnon_coeff}).
This shows that the rapidity spectra strongly constrain the angular distribution of kaon-nucleon scatterings.

\section{Summary and outlook}
\label{sec:summary}
It was shown, within two hadronic transport models, that the differential kaon production cross section in pion-nucleus collisions is sensitive to the decay properties of heavy resonances, and the $K+p$ angular distribution.
In particular, it was found that the transverse momentum spectra provide important constraints for the decay channels producing kaons.
The findings suggest that there exist additional hyperon+$K^*$ decay channels that lead to a decay chain for the kaon, involving additional pions. 
As a result, the momentum of the kaon is reduced in line with the experimental observables.
For larger nuclear targets as tungsten, the final rapidity distribution of the kaons strongly depends on the angular distribution of the kaon scattering cross section with nucleons.
Due to the large size of the effect, an understanding of the anisotropy of the kaon-nucleon interaction in vacuum is necessary, before in-medium effects such as kaon-nucleon potentials (which may also shift the rapidity distribution) can be constrained.

Future works will include studies of the $K^-$ production, for which the mechanisms are very different than for $K^+$, and where spectra, ratios and double ratios were measured by HADES~\cite{Adamczewski-Musch:2018eik}.

\section{Acknowledgements}
\label{sec:ack}
Computational resources have been provided by the Center for Scientific Computing (CSC) at the Goethe-University of Frankfurt and by the Green IT Cube at Gesellschaft für Schwerionenforschung (GSI).
This work was supported by the Helmholtz International
Center for the Facility for Antiproton and Ion Research (HIC for FAIR) within the
framework of the Landes-Offensive zur Entwicklung Wissenschaftlich-Ökonomischer Exzellenz (LOEWE) program launched by the State of Hesse.
V.~S. acknowledges support by the Helmholtz Graduate School for Hadron and Ion Research (HGS-HIRe).
JS thanks the Samson AG and the Bundesministerium für Bildung und Forschung (BMBF) through the ErUM\footnote{Erforschung von Universum und Materie}-Data project for funding. 
\appendix

\bibliography{inspire,non_inspire}

\end{document}